\documentclass[11pt,fullpage]{article}

\usepackage{url,amsmath,amssymb,stmaryrd,proof}

\def\la{\leftarrow}
\def\ra{\rightarrow}
\def\ex{\exists}

\def\bd{\noindent\bf}
\def\sbd{\vspace{8pt}\noindent\bf}

\newbox\tempa
\newbox\tempb
\newdimen\tempc
\def\mud#1{\hfil $\displaystyle{\mathstrut #1}$\hfil}
\def\rig#1{\hfil $\displaystyle{#1}$}
\def\irulehelp#1#2#3{\setbox\tempa=\hbox{$\displaystyle{\mathstrut #2}$}%
		        \setbox\tempb=\vbox{\halign{##\cr
	\mud{#1}\cr
	\noalign{\vskip\the\lineskip}%
	\noalign{\hrule height 0pt}%
	\rig{\vbox to 0pt{\vss\hbox to 0pt{${\; #3}$\hss}\vss}}\cr
	\noalign{\hrule}%
	\noalign{\vskip\the\lineskip}%
	\mud{\copy\tempa}\cr}}%
		      \tempc=\wd\tempb
		      \advance\tempc by \wd\tempa
		      \divide\tempc by 2 }
\def\irule#1#2#3{{\irulehelp{#1}{#2}{#3}%
		     \hbox to \wd\tempa{\hss \box\tempb \hss}}}

\begin{document}
\begin{center}{\Large{\bf A Proof Synthesis Algorithm for a Mathematical 
Vernacular in the Calculus of Constructions}}\\[20pt]
{\bf Gilles Dowek}\\[10pt]
{\bf INRIA}\def\thefootnote{\fnsymbol{footnote}}\footnote[1]{This research was
partly supported by ESPRIT Basic Research Action
``Logical Frameworks''.}
\def\thefootnote{\arabic{footnote}}
\\[15pt]
\end{center}

\section*{Introduction}

In \cite{complete} we have develloped a complete proof synthesis method for 
the the Calculus of Constructions which generalizes Huet's proof synthesis
method for Church Higher Order Logic \cite{Huet72}.
We study in this paper a restriction of the algorithm presented in
\cite{complete} which is always terminating and a complete Vernacular 
\cite{deBruijn87} \cite{deBruijn89} for the 
Calculus of Constructions based on this algorithm. This restriction uses a
second order pattern matching algorithm for the Calculus of Construction 
presented in \cite{matching}
which generalizes Huet's second order pattern matching algorithm for simply
typed $\lambda$-calculus \cite{Huet76} \cite{HueLan}.

A preleminary version of this paper, which presented a complete Vernacular
for a restriction of the Calculus of Construction has been presented in 
\cite{bra}.

\section{An Always Terminating Proof Synthesis Method}

We consider a restriction of the method presented in \cite{complete}.

Let $\Gamma$ be a (non constrained, non quantified context) and $T$ a 
proposition $T = (x_{1}:P_{1}) ... (x_{n}:P_{n})P$ ($P$ atomic).

We consider the quantified context $\Gamma[\ex x:T]$.

If $P$ is a sort.

If $P$ is not a sort then for each $i$ $0 \leq i \leq n$ 
and for each variale $w$ which is either a universal variable $\Gamma$ or an
$x_{i}$, $w:(y_{1}:Q'_{1}) ... (y_{p}:Q'_{p})Q'$ ($Q'$ atomic),

We let:
$$Q_{1} = Q'_{1}$$
$$Q_{2} = [y_{1}:Q'_{1}]Q'_{2}$$
$$...$$
$$Q_{q} = [y_{1}:Q'_{1}] ... [y_{q-1}:Q'_{q-1}]Q'_{q}$$
$$Q = [y_{1}:Q'_{1}] ... [y_{q}:Q'_{q}]Q'$$

$$w:(y_{1}:Q_{1})(y_{2}:(Q_{2}~y_{1})) ... (y_{q}:(Q_{q}~y_{1}~...~y_{q-1}))
(Q~y_{1}~...~y_{q})$$
   
We let:
$$\gamma = [\ex h_{1}:(x_{1}:P_{1}) ... (x_{i}:P_{i})Q_{1};$$
$$\ex h_{2}:(x_{1}:P_{1}) ... (x_{i}:P_{i})(Q_{2}~(h_{1}~x_{1}~...~x_{i}));$$
$$...;$$
$$\ex h_{q}:(x_{1}:P_{1}) ... (x_{i}:P_{i})(Q_{q}~(h_{1}~x_{1}~...~x_{i})~...~
(h_{q-1}~x_{1}~...~x_{i}));$$
$$(x_{1}:P_{1}) ... (x_{i}:P_{i}) 
(Q~(h_{1}~x_{1}~...~x_{i})~...~(h_{q}~x_{1}~...~x_{i})) =
(x_{1}:P_{1}) ... (x_{n}:P_{n})P]$$
we consider the subsitutions:
$$\{<x,\gamma, 
[x_{1}:P_{1}] ... [x_{i}:P_{i}]
(w~(h_{1}~x_{1}~...~x_{i})~...~(h_{q}~x_{1}~...~x_{n}))>\}$$

We get the context $\Gamma \gamma$.

Then form letf to right, for each $j$ such that $Q_{j}$ is not second order in
$\Gamma[\ex y_{1}:Q_{1}; ...; \ex y_{j-1}:Q_{j-1}]$ we try to instanciate 
$h_{j}$ by a universal variable. The accouting equation is always a
second-order-argument-restricted problem, we solve it.

Then the equation of the context is a second-order-argument-restricted
problem we solve it.

At last from right to left, we try to instanciate the existential variables
of the context by a universal variable or a term of the form
$[x_{1}:P_{1}] ... [x_{i}:P_{i}]x_{i}$. Accounting equations are
second-order-argument-restricted problem, we solve them.

{\sbd Proposition:} This algorithm is always terminating

{\sbd Definition}: Transitive Closure of a Proof Synthesis Method

We consider a proof synthesis method.
We write $\Gamma \leadsto P$ the assertion that in the context 
$\Gamma$ a proof of $P$ is synthetized by the method.
We consider also assertions $\Gamma \hookrightarrow P$ meaning intuitively 
that there exists a text in Vernacular which is a demonstration of $P$.

We want, a priori, to have only one rule that allows to synthesize the proof 
of a new proposition, using already proved ones. 
Actually we need also another rule, which allows to introduce explicitly an
hypothesis or a variable. Indeed, let us imagine that we want to prove a 
proposition
$A \ra B$ in introducing the hypothesis $A$ then proving $B$. If a proof of
$B$ cannot be automatically synthesized and for instance we have to prove a
lemma $C$ (using the hypothesis $A$) before, we cannot let the system
introduce automatically the hypothesis $A$, we have to do it by hand.

{\sbd Rule 1: Synthesis}

$$\irule{\Gamma \hookrightarrow Q_{1}~~...~~\Gamma \hookrightarrow Q_{n}~~~~
            \Gamma[c_{1}:Q_{1}; ...; c_{n}:Q_{n}] \leadsto P}
           {\Gamma \hookrightarrow P}
           {c_{i}~\mbox{not free in}~P}$$

{\sbd Rule 2: Explicit introduction}

$$\irule{\Gamma[x:Q] \hookrightarrow P}
           {\Gamma \hookrightarrow (x:Q)P}
           {}$$

{\sbd Lemma:} If a proof synthesis method is sound then so is its transitive 
closure.

{\bd Proof:} 
Let $\Gamma$ a context and $P$ a proposition such that we know a derivation 
of $\Gamma \hookrightarrow P$. By induction on the length of this derivation, 
we construct a proof $t$ of $P$  in $\Gamma$.

If the last rule of the derivation is the rule {\it Synthesis} we have by the
soundness of the proof synthesis method a term $u$ such that:
$$\Gamma[c_{1}:Q_{1}, ..., c_{n}:Q_{n}] \vdash u:P$$ 
and by induction hypothesis terms $v_{1}, ..., v_{n}$ such that:
$$\Gamma \vdash v_{i}:Q_{i}$$
so:
$$\Gamma \vdash u[c_{1} \la v_{1}, ..., c_{n} \la v_{n}]:P$$

If the last rule of the derivation is the rule {\it Explicit introduction}
then $P = (x:U)V$ and we have by induction hypothesis a term $u$ such that:
$$\Gamma[x:U] \vdash u:V$$
$$\Gamma \vdash [x:U]u:P$$

{\sbd Defintion:} Transitively Complete Proof Synthesis Method

A proof synthesis method is said to be {\it transitively complete} if for 
all context $\Gamma$ and proposition $P$  if there exists a term $t$ such that
$\Gamma \vdash t:P$, then $\Gamma \hookrightarrow P$.

{\sbd Proposition:} The proof synthesis method presented above if
transitively complete.

{\sbd Premises:}

\section{An Allusive Vernacular}

In the Elementary Vernacular \cite{namingscoping} there is an instruction 
\verb%Proof t.% where $t$ is a term. When the proof checker meets
the instruction  \verb%Proof t.% in a context $\Gamma$, it computes the type
of $t$ in $\Gamma$, eliminates the local elements declared since the last
instruction \verb%Statement%, checks that the type obtained that way is
the same as the one given in the last instruction \verb%Statement% 
(if it is not then it fails),
then eliminates the local elements declared since the last
instruction \verb%Theorem% and adds this new theorem to the context.

We modify this Vernacular in replacing this instruction by 
\verb%Using s1, ..., sn.% where $s1, ..., sn$ are symbolic names.
When the proof checker meets the instruction \verb%Using s1, ..., sn.% in a 
context $\Gamma$, it computes, using the type given in the last instruction 
\verb%Statement% and 
the local elements declared since this last instruction, the goal to be proved
then looks for a proof of this goal using the premises $\{s1, ...,
sn\}$, fails if it does not find one, then eliminates the local elements
declared since the last
instruction \verb%Theorem% and adds this new theorem to the context.

{\sbd Proposition:}
For all inhabited proposition $P$, there exists a text in Vernacular which
denotes a proof of this proposition.

{\bd Proof:} 
We construct such a text from a derivation of $\Gamma \hookrightarrow P$.
\begin{itemize}
\item{
If the last rule used is the rule {\it Synthesis}, then there exists, by
induction hypothesis, texts that define symbols $c_{1}, ..., c_{n}$ proof 
of $Q_{1},...,Q_{n}$. Let \verb%<text>% be their concatenation.

There exists also a set of premises \verb%<premises>% of 
$\Gamma[c_{1}:Q_{1}, ..., c_{n}:Q_{n}]$ used in the synthesis of a proof of
$P$. We build the text:
\begin{verbatim}
Remark <name>.
Statement P.
   <text>
Using <premises>.
\end{verbatim}
     }
\item{
If the last rule is the {\it Explicit introduction} then $P = (x:U)V$.
By induction hypothesis $V$ has a proof in Vernacular in the context
$\Gamma[x:U]$ that uses a set \verb%<premises>% of premises
($x$ may belong to \verb%<premises>%). Let a text in Vernacular that proves 
$V$ in this context:
\begin{verbatim}
Remark <name>.
   <text 1>
Statement <statement>.
   <text 2>
Using <premises>.
\end{verbatim}
We transform this text in:
\begin{verbatim}
Remark <name>.
   Variable/Hypothesis x:U.
   <text 1>
Statement <statement>.
   <text 2>
Using <premises>.
\end{verbatim}
     }
\end{itemize}

\end{document}